\documentclass[aps,preprint]{revtex4-1}

\usepackage{graphicx}
\usepackage{dcolumn}
\usepackage{bm}
\usepackage{xcolor}
\usepackage{amssymb}   
\usepackage{amsmath}
\usepackage{ulem}
\usepackage{times}

\begin{document}
\title{
Half-integer quantized anomalous thermal Hall effect in the Kitaev material {\boldmath $\alpha$}-RuCl{\boldmath $_3$}
}

\author{T. Yokoi$^{1}$}
\altaffiliation{These authors contributed equally to this work.}
\author{S. Ma$^{1,\ast}$}
\author{Y. Kasahara$^1$}
\author{S. Kasahara$^1$}
\author{T. Shibauchi$^2$}
\author{N.~Kurita$^3$}
\author{H. Tanaka$^3$}
\author{J. Nasu$^4$}
\author{Y. Motome$^5$}
\author{C. Hickey$^6$}
\author{S. Trebst$^6$}
\author{Y. Matsuda$^1$}
\email{matsuda@scphys.kyoto-u.ac.jp}

\affiliation{$^1$Department of Physics, Kyoto University, Kyoto 606-8502, Japan}
\affiliation{$^2$Department of Advanced Materials Science, University of Tokyo, Chiba 277-8561, Japan}
\affiliation{$^3$Department of Physics, Tokyo Institute of Technology, Meguro, Tokyo 152-8551, Japan}
\affiliation{$^4$Department of Physics, Yokohama National University, Hodogaya, Yokohama 240-8501, Japan}
\affiliation{$^5$Department of Applied Physics, University of Tokyo, Bunkyo, Tokyo 113-8656, Japan}
\affiliation{$^6$Institute for Theoretical Physics, University of Cologne, 50937 Cologne, Germany}

\begin{abstract}
%

Heat transport mediated by Majorana edge modes in a magnetic insulator leads to a half-integer thermal quantum Hall conductance, which has recently been reported for the two-dimensional honeycomb material $\alpha$-RuCl$_3$. 
While the conventional electronic Hall effect requires a perpendicular magnetic field, we find that this is not the case 
in $\alpha$-RuCl$_3$. Strikingly, the thermal Hall plateau appears even for a magnetic field with no out-of-plane components.   
The field-angular variation of the quantized thermal Hall conductance has the same sign structure of the topological Chern number, 
which is either $\pm$1, as the Majorana band structure of the pure Kitaev spin liquid.   
This observation of a half-integer anomalous thermal Hall effect firmly establishes that the Kitaev interaction is primarily responsible and that the non-Abelian topological order associated with fractionalization of the local magnetic moments persists even in the presence of non-Kitaev interactions in $\alpha$-RuCl$_3$.

\end{abstract}

\maketitle
The discovery of the integer quantum Hall effect (QHE) in a two-dimensional (2D) electron gas has opened a window for exploring topological phases of quantum matter.   
When topology is combined with strong 
correlations, even more interesting phenomena can arise, such as the emergence of collective excitations having fractional quantum numbers with respect to the elementary particles. 
One of the most prominent examples is the fractional QHE, in which the constituent particles are electrons but the quasiparticles 
carry fractions of the electron charge.  
Another venue for the interplay of topology and correlations 
are quantum spin liquids in magnetic insulators \cite{Balents10}, 
where fluctuations prevent the formation of magnetic order down to $T$=\,0 and instead lead to the emergence of macroscopic
entanglement.
%
%
A paradigmatic example is the Kitaev spin liquid \cite{Kitaev06}, which forms for strongly interacting spins in a magnetic field
and exhibits anyonic quasiparticles in the form of massive $Z_2$ vortices that bind a Majorana zero mode.  
The non-Abelian character of these Ising anyons in the Kitaev spin liquid is identical to that of certain topological superconductors 
\cite{Read00}
 and the $\nu=5/2$ fractional quantum Hall state \cite{MooreRead91}.


Direct experimental evidence for the formation of a topological state with non-Abelian anyons has long been sought-after. 
In fact, it required the observation of yet another QHE -- a 
quantized {\it thermal} Hall effect,
reported last year in experiments for both the $\nu=5/2$ fractional quantum Hall state \cite{Banerjee18}
and the spin-orbit entangled Mott insulator $\alpha$-RuCl$_3$ \cite{Kasahara}.
In both experiments the transverse thermal Hall conductivity (per 2D sheet) $\kappa_{xy}^{\rm 2D}$ exhibits a quantized plateau 
at half integer values,
\begin{equation}
	\kappa_{xy}^{\rm 2D}=\frac{1}{2} C_h \cdot K_0 \,, 
\end{equation}
where $K_0=(\pi^2/3)(k_B^2/h)T$ is the quantum thermal conductance, in analogy to the quantum electronic conductance $2e^2/h$
for the electronic QHE, and $C_h$ is an integer related to the number of chiral edge modes ($C_h=1$ for $\alpha$-RuCl$_3$ and $C_h=5$ for the $\nu=5/2$ state).
The half-integer quantization is, in both cases, direct evidence of charge neutral Majorana fermions carrying heat along the sample edges, and as such direct evidence for the non-Abelian topological nature of the bulk state. Here, we show that the fundamental nature of the QHE in the two systems, however, must be clearly distinguished. For the two-dimensional electron gas, the QHE arises from the electrons experiencing a Lorentz force and the formation of {\it Landau levels} in a perpendicular magnetic field and the subsequent splitting of the degenerate states of partially filled Landau levels through Coulomb interactions. In this case, $C_h$ is related to the number of upstream/downstream edge modes and its sign is determined purely by the magnetic field perpendicular to the sample. 
In the insulator $\alpha$-RuCl$_3$, we show that a different physical mechanism must be at play, consistent with a Kitaev spin liquid (KSL) that
forms in the presence of a magnetic field. In this scenario the local magnetic moments interact via bond-directional Ising-type 
interactions, which at low temperatures leads to a fractionalization into itinerant Majorana fermions and static $Z_2$ vortices 
located on the hexagons of the honeycomb lattice. In the absence of a magnetic field, the Majorana fermions form 
a semi-metal with a Dirac cone dispersion. Applying a magnetic field, the Majorana spectrum gaps out into 
 a {\it Chern insulator}, i.e.~a band insulator whose bands are endowed with a non-trivial Berry curvature, with the resulting 
Chern number $\pm 1$ for each band. The non-trivial topology of this time-reversal symmetry breaking insulator 
also gives rise to chiral Majorana modes at the edges, with the Chern number identified with $C_h$ above, and thus a half-integer quantized thermal QHE with the sign structure of 
$\kappa_{xy}^{\rm 2D}$ reflecting the non-trivial bulk topology. 


Here, we report the observation of an {\it anomalous} thermal QHE in $\alpha$-RuCl$_3$ that gives rise to a half-integer quantized thermal Hall
conductance plateau even in the {\it absence} of an out-of-plane magnetic field, thereby confirming that the thermal QHE in 
$\alpha$-RuCl$_3$ does indeed arise from a topologically non-trivial Chern insulator of Majorana fermions. 
Our experiments build on recent work identifying the spin-orbit coupled Mott insulator $\alpha$-RuCl$_3$ \cite{Plumb14}
as a Kitaev material \cite{Jackeli09,Trebst17,Takagi19}, in which local $j_{\rm eff}=1/2$ pseudospins are almost coplanar within the honeycomb lattice \cite{Kim15}, see Fig.\,1A.
Significant bond-directional Kitaev interactions $J_K/k_B\approx 80$\,K have been reported \cite{Rau14,Winter17}, which dominate over all other non-Kitaev interactions (such as Heisenberg and off-diagonal exchange). The latter induce antiferromagnetic (AFM) ordering with a zigzag magnetic structure  at $T_N\approx7.5$\,K \cite{Johnson15}. A double peak structure in the temperature dependence of the specific heat \cite{Do17,Widmann19}, fermionic excitations in Raman scattering spectra \cite{Sandilands15,Nasu16}, and a broad energy continuum in inelastic neutron scattering \cite{Do17,Banerjee16,Yoshitake16,Banerjee17,Balz19} above $T_N$ have been interpreted as signatures of fractionalization. 

The AFM order in $\alpha$-RuCl$_3$ is strongly suppressed by applying an in-plane magnetic field  {\boldmath $H$} parallel to the 2D honeycomb planes, leading to a 
field-induced paramagnetic state (Fig.\,1B).  
To probe the emergence of fractional quasiparticles in this regime, thermal transport measurements are a powerful technique \cite{Nasu17}, as the emergent charge neutral Majorana fermions do not respond to electric fields but can carry heat via a subtle
coupling to the phonons \cite{Rosch18,Balents18}.  
Such non-trivial magnetic excitations can be detected, in particular, via thermal Hall conductivity $\kappa_{xy}$ measurements \cite{Kasahara18,Hentrich19} transverse to a heat current {\boldmath $q$}.  Very recently, a thermal Hall plateau quantized at a half-integer value of $\kappa_{xy}$ has been observed \cite{Kasahara} for a field {tilted} away from the crystallographic $c$ axis, 
proving the spin liquid nature of the field-induced phase.  
The observation of the half-integer thermal QHE immediately raises several fundamental questions about the nature of the spin liquid state.  In fact, the  thermal QHE is observed at {\it intermediate} magnetic field strengths and in proximity to a magnetically ordered state -- an experimental scenario that cannot be captured by the pure Kitaev model, in which the KSL originates in the small-field limit from a gapless spin liquid.
Moreover, the influence of non-Kitaev interactions, which are inevitably present in $\alpha$-RuCl$_3$ \cite{Rau14,Winter17}, on the KSL has been an issue under intense debate. 
To demonstrate the unaltered nature of the observed spin liquid state at intermediate field strengths from the theoretically well understood KSL, we here determine the field-angular variation of the thermal conductance plateau, which provides stringent experimental evidence for a Chern insulator that closely mimics the angular dependence of the KSL.

We measured  $\kappa_{xy}$ of five single crystals from the same batch with {\boldmath $q$}$\parallel a$ in  {\boldmath $H$} applied to a tilted direction in the $ac$ plane with $\theta=-60^{\circ}$, where $\theta$ is the polar angle measured from the $c$ axis (Figs.\,S1 A-C).   The quantized thermal Hall plateau at one half of $K_0$  was observed in three crystals (\#1$-$\#3), 
while it was absent in one crystal (\#4) with low longitudinal thermal conductivity $\kappa_{xx}$ 
and in one crystal (\#5), which shows a partial magnetic order at 14\,K due to stacking faults.  
These results 
indicate that high-quality single crystals are necessary for the half-integer thermal QHE.  
The magnetic field strengths at which the quantization occurs is slightly sample dependent.   
Here crystal \#3  is used for the measurements.  Figure\,1B shows the phase diagram of $\alpha$-RuCl$_3$ in {\boldmath$H$} applied parallel to the $a$ axis (zigzag direction of a honeycomb lattice).  A critical magnetic field $H_{AF}$, at which AFM order disappears,  is determined by the minimum of  $\kappa_{xx}(H)$ (Fig.\,S2). The determined $H_{AF}$ is close to those determined by other methods \cite{Balz19} but is smaller than that reported by magnetic torque measurements \cite{Modic19}.     
 
Varying the field direction, we depict the $H$-dependence of $\kappa_{xy}$ at 4.8\,K ($\mu_0H_{AF}\approx 6.5$\,T) for {\boldmath $H$} applied antiparallel to the $a$ axis by the red circles in Figs.\,2A and 2B.
What is remarkable is that, even for an in-plane magnetic field with {\it no} out-of-plane component,
 a finite $\kappa_{xy}$ with positive sign is observed in the spin liquid regime (left figure of Fig.\,2C).    In the AFM phase, $\kappa_{xy}$ is absent below $\sim 5$\,T within the resolution and  increases with approaching $H_{AF}$.    Upon entering the spin liquid state, $\kappa_{xy}$  shows an upward increase. On the right axes of Figs.\,2A and 2B,  thermal Hall conductance per 2D layer,  $\kappa_{xy}^{\rm 2D}=\kappa_{xy}d$, where $d$=0.57\,nm is the interlayer distance, is plotted in units of $K_0$.   Taking into account the ambiguity of measuring the contact distances, $\kappa_{xy}^{\rm 2D}$ exhibits a quantized plateau at one half of $K_0$, within an error bar of  $\pm$ 10\%, above $\sim$9.7\,T (Fig.\,2B).  Above $\sim 11.5$\,T,  $\kappa_{xy}^{\rm 2D}$ is strongly suppressed from the half quantized value.   Weakly $H$-dependent $\kappa_{xy}$ between $\mu_0H$=7 and 8.5\,T is unrelated to the topological property, because  $\kappa^{xy}/T$ in this regime is temperature dependent (Fig.\,S3).  On the other hand,  $\kappa_{xy}$ is strongly suppressed or vanishes when {\boldmath $H$} is applied parallel to the $b$ axis (armchair direction of a honeycomb lattice)  and a half-integer thermal QHE is not observed as shown by blue circles in Figs.\,2A and 2B (right figure of Fig.\,2C).    The absence of the transverse signal for $\bm{H}\parallel b$ is expected from the crystal structure, which possesses a two-fold symmetry around the $b$ axis.  In contrast, no two-fold symmetry exists around the $a$ axis,  because the Ru sites form two sublattices. 

A rapid reduction of $\kappa_{xy}^{\rm 2D}$ at low temperatures, followed by a half quantized plateau similar to the present {\boldmath $H$}$\parallel -a$ results, has also been reported in previous measurements \cite{Kasahara} for {\boldmath $H$} applied in a {\it tilted} direction in the $ac$ plane with $\theta=-60^{\circ}$ and $-45^{\circ}$.  
Figure\,2D and its inset show $T$-dependence of $\kappa_{xy}/T$ at $\mu_0H=10.5$\,T where the half-integer thermal QHE is observed for the in-plane field.   The thermal QHE persists up to $\sim$5\,K, above which $\kappa_{xy}/T$ is enhanced from the quantized value, peaks at around 15\,K and decreases.  A similar $T$-dependence has also been reported for $\theta=-45^{\circ}$ and $-60^{\circ}$ \cite{Kasahara}.  Thus the thermal Hall response in the spin liquid state for {\boldmath $H$}$\parallel -a$ is essentially the same as that for  $\theta=-45^{\circ}$ and $-60^{\circ}$ including the Hall sign.  The enhancement from the quantized value at high temperature has been discussed in terms of the signature of a topological phase transition \cite{Moon19}.
What sets the current experiments apart is that the  observation of a half-integer thermal QHE with no out-of-plane field unambiguously demonstrates that the quantization effect is unrelated to the Lorentz force or the formation of Landau levels, but in fact the manifestation of an anomalous thermal QHE originating from the formation of
a Chern insulator of Majorana fermions.

Further varying the field direction, Figs.\,3A and 3B depict $\kappa_{xy}$ at $\theta=\pm60^{\circ}$ at 4.3\,K.   Finite $\kappa_{xy}$ is observed both in the AFM and spin liquid regimes.   The half-integer quantized thermal Hall conductance with positive sign followed by the rapid reduction at high field is observed at $\theta=-60^{\circ}$, as reported previously.   The half-integer quantized thermal Hall conductance is also observed at $\theta=+60^{\circ}$ but, strikingly, its sign is negative in the spin liquid state.  Moreover, this sign change vanishes in the AFM phase below 6\,T for which $\kappa_{xy}$ at $\theta=\pm60^{\circ}$ coincide within the experimental resolution.   We note that the positive $\kappa_{xy}$ for $\theta=-90^{\circ}$ ($\bm{H}\parallel -a$, see Fig. 2) means, as required by crystal symmetry,  a negative  $\kappa_{xy}$ for $\theta=90^{\circ}$.   
Taken together these out-of-plane results demonstrate a {\it varied} non-trivial sign structure of  $\kappa_{xy}$ in the spin liquid regime with respect to the angle $\theta$ of the $c$ axis within the $ac$ plane.  
In sharp contrast, such a sign change for $\theta=\pm60^\circ$ is absent well inside the AFM phase, demonstrating that the mechanism behind the thermal Hall effect in the spin liquid state is fundamentally different from that in the AFM phase. 
The finite $\kappa_{xy}$ in the AFM phase might be attributed to a magnon thermal Hall effect (arising from the Berry curvature of magnon bands \cite{Onose10}).  As the magnons are bosonic quasiparticles, $\kappa_{xy}^{\rm 2D}$ is not quantized \cite{Cookmeyer18}.    
{The absence of a sign change of $\kappa_{xy}$ between $\theta=\pm 60^{\circ}$ and the absence of $\kappa_{xy}$ for {\boldmath $H$}$\parallel a$  well inside the AFM phase imply that such a magnon thermal Hall effect is dominated by a finite out-of-plane magnetic field. }
The eventual splitting of $\kappa_{xy}$ in the AFM phase upon approaching the transition at $H_{AF}$ for both $\theta=-90^{\circ}$ (green area in Fig.\,1B) and $-60^{\circ}$ is likely due to fluctuation effects in the proximity of the phase transition.  


Both the half-integer thermal QHE without an out-of-plane magnetic field and the nontrivial sign change of $\kappa_{xy}$ with respect to the $c$ axis are in stark contrast to the conventional Hall effect of the 2D electron gas, but can be naturally explained for the KSL. In the pure Kitaev model, adding a weak Zeeman field generates, at third order in perturbation theory, a three-spin interaction term $\!\propto \!\! -\frac{h_xh_yh_z}{J^2}\sum_{\langle j,k,\ell \rangle}\sigma_j^x\sigma_k^y\sigma_{\ell}^z$ (where ${\langle j,k,\ell \rangle}$ stands for three neighboring sites). Crucially, this term breaks time reversal symmetry and opens up a Majorana gap in the bulk. It manifests as a next nearest neighbor hopping of Majorana fermions and gives rise to a non-trivial topological band structure, with an associated topologically protected chiral Majorana edge current \cite{Kitaev06}. The Chern number $C_h$ of the gapped Majoranas can be determined as
\begin{equation}
C_h={\rm sgn}(h_xh_yh_z) \,,
\end{equation}
where $h_x,h_y$ and $h_z$ are the field components with respect to the spin axes $S^x,S^y$ and $S^z$ (Fig.\,1A), respectively. We note that the directions of the spin axes ($x$, $y$ and $z$) are determined by both the spin-orbit interaction and the crystal structure, and are different from the directions of the crystal axes ($a, b$, and $c$). Experimentally, $C_h$ can be readily determined by the sign of $\kappa_{xy}^{\rm 2D}$ in the half-integer thermal QHE regime. A measurement of the field-angular dependence of the half-integer thermal QHE thus provides crucial information on the topology of the Majorana fermion band structure, and a crucial test of whether 
the Kitaev interaction is responsible for the thermal QHE.

The field-angular variation of the Chern number for the pure Kitaev model, with respect to the spin and crystal axes, is shown in Figs.\,4A-C. 
Note that $C_h$ is finite even in the absence of an out-of-plane magnetic field; e.g.~$C_h=-$1(+1) for {\boldmath $H$} parallel (antiparallel) to the $a$ axis (see purple arrow in Fig.\,4{\bf C}).  When the applied magnetic field has no out-of-plane component, one might suspect that the net chirality of any resultant edge states cannot be determined uniquely.  However, this is only true when the field is applied along a high-symmetry axis with a two-fold rotation symmetry. Such a symmetry would reverse the chirality of the edge states, ensuring their absence when the symmetry is present. 
This guarantees that the thermal conductance vanishes for a field along the $b$ axis  (see dashed purple arrow in Fig.4{\bf C}), which does possess a two-fold rotation symmetry, while there is a finite response for a field along the $a$ axis, which does not (a two-fold rotation would swap the A and B sublattices, as shown by black and white circles in Fig.\,4D).  The small but finite $\kappa_{xy}$ experimentally observed for {\boldmath $H$}$\parallel b$ is likely attributed to a misalignment of the magnetic field direction from the $b$ axis.

In the pure Kitaev model, when the magnetic field is rotated within the $ac$ plane $C_h$ changes its sign at $\theta\approx 35^{\circ}$ and $C_h$=1 and $-1$ for $\theta=-60^{\circ}$ and $\theta=60^{\circ}$, respectively (see dashed orange and orange arrows in Fig.\,4{\bf B}). For a real material, it is natural to wonder whether this sign change can persist in the presence of non-Kitaev interactions. Generically such interactions will give rise to a {\it linear} in field contribution to the Chern number as $C_h={\rm sgn}\{c_1(h_x+h_y+h_z)+c_3h_xh_yh_z+\dots\}=\pm1$. For in-plane fields, the sum $(h_x+h_y+h_z)$ vanishes, leaving the third order Kitaev contribution as the leading term. As a result, any additional interactions are not expected to play a significant role. For out-of-plane fields however the linear contribution survives and could in principle dramatically alter the sign structure of the Chern number, and hence the thermal Hall conductance. The fact that a sign change of $\kappa_{xy}$ between $\pm 60^{\circ}$ is experimentally observed places constraints (\cite{SM}, section 2) on the ratio of $c_1/c_3 \lesssim 0.2 J_K^2$. We thus conclude that the contribution of non-Kitaev interactions to the observed half-quantized thermal QHE is vastly outnumbered by those of the Kitaev exchange.

The sign structure of the observed half-integer thermal QHE clearly shows that the non-trivial topology of the Majorana fermion bands in $\alpha$-RuCl$_3$ is consistent with that of the pure Kitaev spin liquid, identifying $\alpha$-RuCl$_3$ as the host of a new quantum phase of matter -- a Chern insulator of neutral Majorana fermions.  The present results demonstrate that the Kitaev interaction is responsible for the quantization of the thermal Hall effect and that the expected non-Abelian topological order persists even in the presence of non-Kitaev interactions.

	\bigskip
	\bigskip
	\noindent
	{\bf Acknowledgements}
	
	We thank Y. Akagi, S. Fujimoto, H.-Y. Kee,  and E.-G. Moon for useful discussions.  {\bf Funding:} This work was supported by Grants-in-Aid for Scientific Research (KAKENHI) (numbers 15H02106, 15K13533, 16H02206, 16H00987, 16K05414, 17H01142, 18H01177, 18H01180, 18H04223, 18H05227, and 19K03711), Grants-in-Aid for Scientific Research on innovative areas ``Topological Materials Science" (number JP15H05852) from Japan Society for the Promotion of Science (JSPS), and JST CREST (JP-MJCR18T2). 
	J.N. acknowledge the support of Leading Initiative for Excellent Young Researchers in MEXT. 
	C.H. and S.T. acknowledge funding by the Deutsche Forschungsgemeinschaft (DFG, German Research Foundation) -- project numbers 277101999 and 277146847 -- TRR 183 (project B01) and CRC 1238 (project C03).\\
	
	\bigskip
	\noindent
	{\bf Author contributions:}
	Y.K. and Y. Matsuda conceived and designed the study. T.Y., S.M. and  Y.K. performed the thermal transport measurements. T.Y., S.M., Y.K., S.K, and Y. Matsuda analyzed the data. N.K. and H.T. synthesized the high-quality single crystalline samples. Y.K., T.S., J.N., Y. Motome, C.H., S.T. and Y. Matsuda prepared the manuscript.
	
	\bigskip
	\noindent
	{\bf Competing interests:}
	The authors declare no competing interests. 
	
	\bigskip
	\noindent
	{\bf Data and materials availability:}
	All the data in the manuscript main text are available as .csv ASCII files in the supplementary materials.

	
	%

\newpage

\begin{figure}[h!]
	\begin{center}
	\hspace*{-30mm}
		\includegraphics[width=1.0\linewidth]{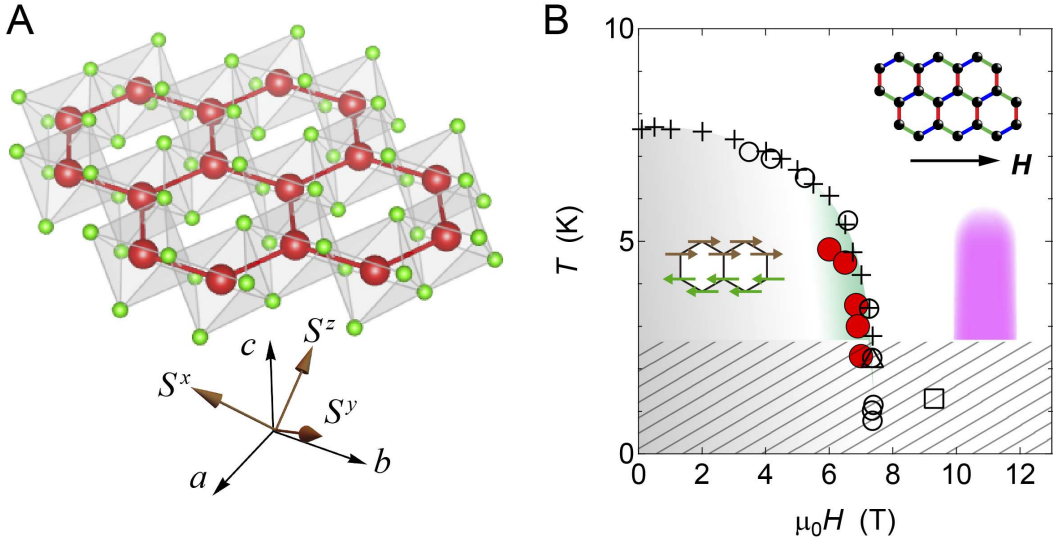}
		\caption{ 
			({\bf A}) Upper figure illustrates the crystal structure of $\alpha$-RuCl$_3$ in the $ab$ plane.  $\alpha$-RuCl$_3$ is composed of very weakly coupled 2D honeycomb layers of edge-sharing RuCl$_3$ octahedra. Red and green circles represent Ru [Ru$^{3+}(4d^5)$, $j_\mathrm{eff}=1/2$] and Cl$^{-}$ ions.   $a$ ($b$ ) axis corresponds to zigzag (armchair) direction in a honeycomb lattice.  The lower figure shows spin $x$, $y$ and $z$ axes (brown arrows), which are determined both by the spin-orbit interaction and crystal structure,  crystal $a$, $b$ and $c$ axes (black arrows).  Crystal $a$, $b$ and $c$ directions correspond to  $(1,-1,2)$, $(1,-1,0)$ and $(1,1,1)$ directions in the spin axis, respectively.  ({\bf B}) $H$-$T$ phase diagram in magnetic field applied parallel to the $a$ axis.  Pink area represents the regime where half-integer thermal QHE is observed. The gray shaded area represents the region of zigzag antiferromagnetic order, where $\kappa_{xy}$ strongly suppressed or vanishes.  The green shaded area represents the regime where finite $\kappa_{xy}$ appears.  Red circles indicate antiferromagnetic transition field  $H_{AF}$ determined by the longitudinal thermal conductivity (Fig.\,S2).  Crosses, open circles,  triangle and square indicate $H_{AF}$ determined by magnetic susceptibility, magnetocaloric effect, neutron elastic scattering \cite{Balz19} and magnetic torque \cite{Modic19}, respectively.  Hatched area is the region where thermal Hall effect could not be measured. 
		}
	\end{center}
\end{figure}

\newpage
\begin{figure}[h!]
	\begin{center}
		\includegraphics[width=1.0\linewidth]{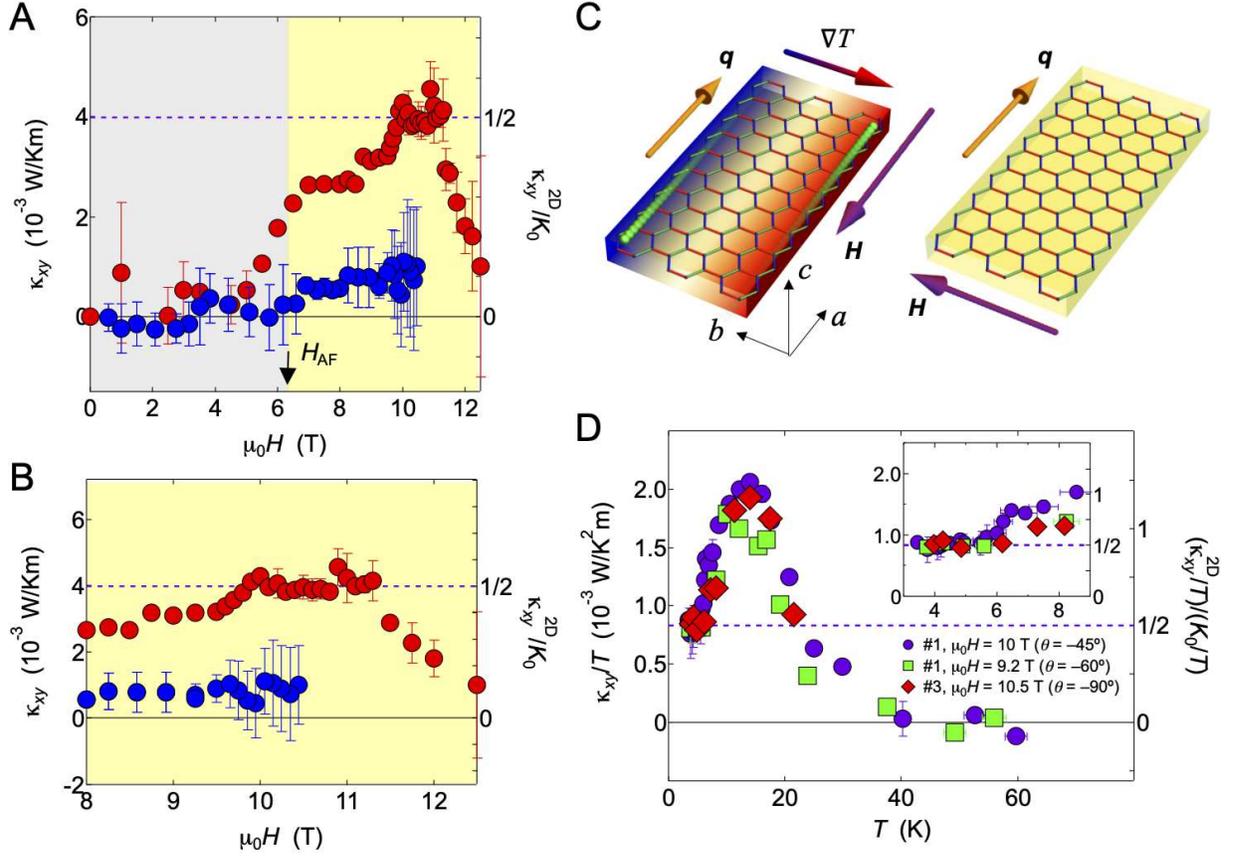}
		\caption{
	({\bf A}) Thermal Hall conductivity $\kappa_{xy}$ in the antiferromagnetic (gray shaded area) and spin liquid (yellow) states for {\boldmath $H$}$\parallel -a$ axis (red circles) and  {\boldmath $H$}$\parallel b$ axis (blue circles) at 4.8\,K.  In the right axis,  thermal Hall conductance per 2D layer $\kappa_{xy}^{\rm 2D}$  is plotted in units of  the quantum thermal conductance $K_0=(\pi^2/3)(k_B^2/h)T$.  The small but finite $\kappa_{xy}$ experimentally observed for {\boldmath $H$}$\parallel b$ is likely attributed to a misalignment of the magnetic field direction from the $b$ axis. ({\bf B}) The same data in high field spin liquid region.    ({\bf C}) Schematics of the transverse temperature gradient generated by thermal current {\boldmath $q$} applied along $a$ axis.  Left: When {\boldmath $H$} is applied along $-a$ axis, temperature gradient $\nabla T$ appears along $b$ axis.  The red (blue) area indicates high (low) temperature region.  Green circles represent the edge current of itinerant Majorana particles. Right: When  {\boldmath $H$} is applied along $b$ axis, no transverse temperature gradient appears along the $b$ axis.   ({\bf D})  Temperature dependence of $\kappa_{xy}/T$ for {\boldmath $H$}$\parallel -a$ at the magnetic field where the thermal QHE is observed.  The right axis shows $\kappa_{xy}^{\rm 2D}/T$  normalized by $K_0/T$.  The results for $\theta=-60^{\circ}$ and  $-45^{\circ}$ of different crystal reported previously are also plotted.  The inset shows the same data at low temperature range. 
		}
	\end{center}
\end{figure}
\newpage
\begin{figure}[h!]
	\begin{center}
		\includegraphics[width=1.0\linewidth]{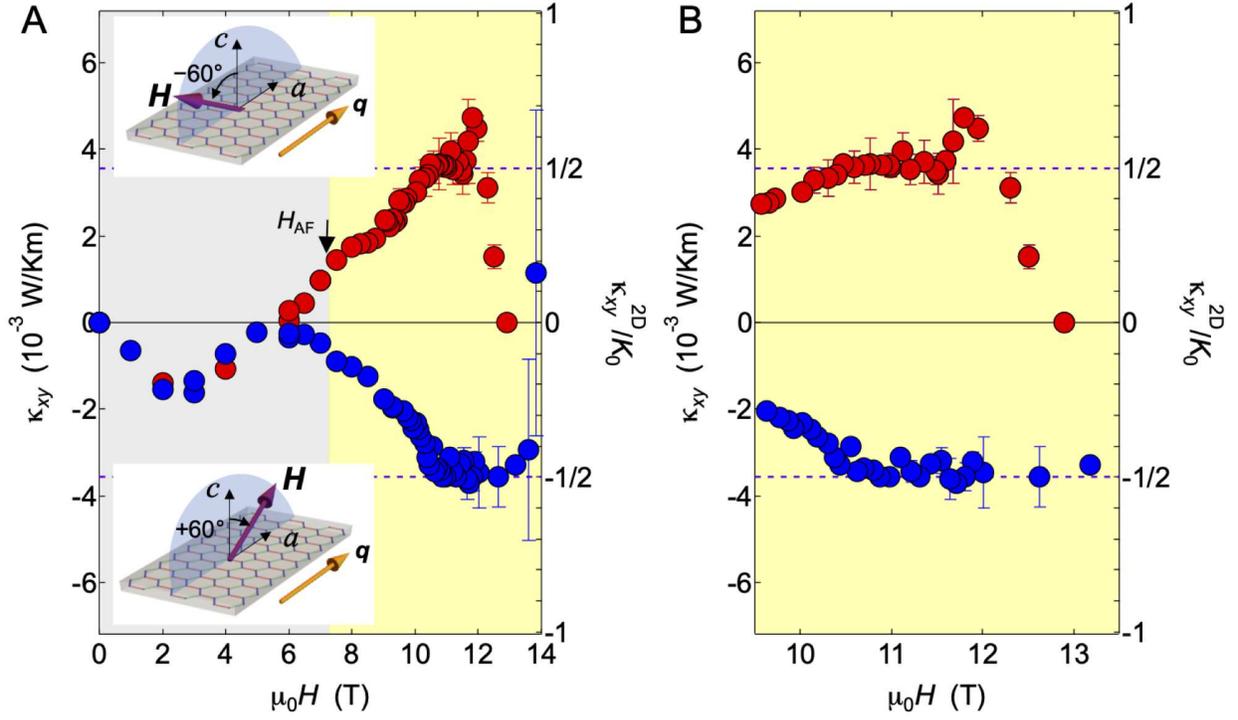}
		\caption{({\bf A}) Field dependence of $\kappa_{xy}$ at 4.3\,K in tilted field of  $\theta=-60^{\circ}$ (red circles) and 60$^{\circ}$ (blue circles) away  from the $c$ axis in the $ac$ plane.  Gray and yellow shaded areas represent the AFM ordered and spin liquid states, respectively.   Thermal current {\boldmath $q$} is applied along the $a$ axis.   The antiferromagnetic transition field determined by the minimum of $\kappa_{xx}(H)$  is $H_{AF}$=7.0\,T ({\bf B}) The same data in the high field spin liquid state.  	
		}
	\end{center}
\end{figure}
\newpage
\begin{figure}[h!]
	\begin{center}
		\includegraphics[width=0.8\linewidth]{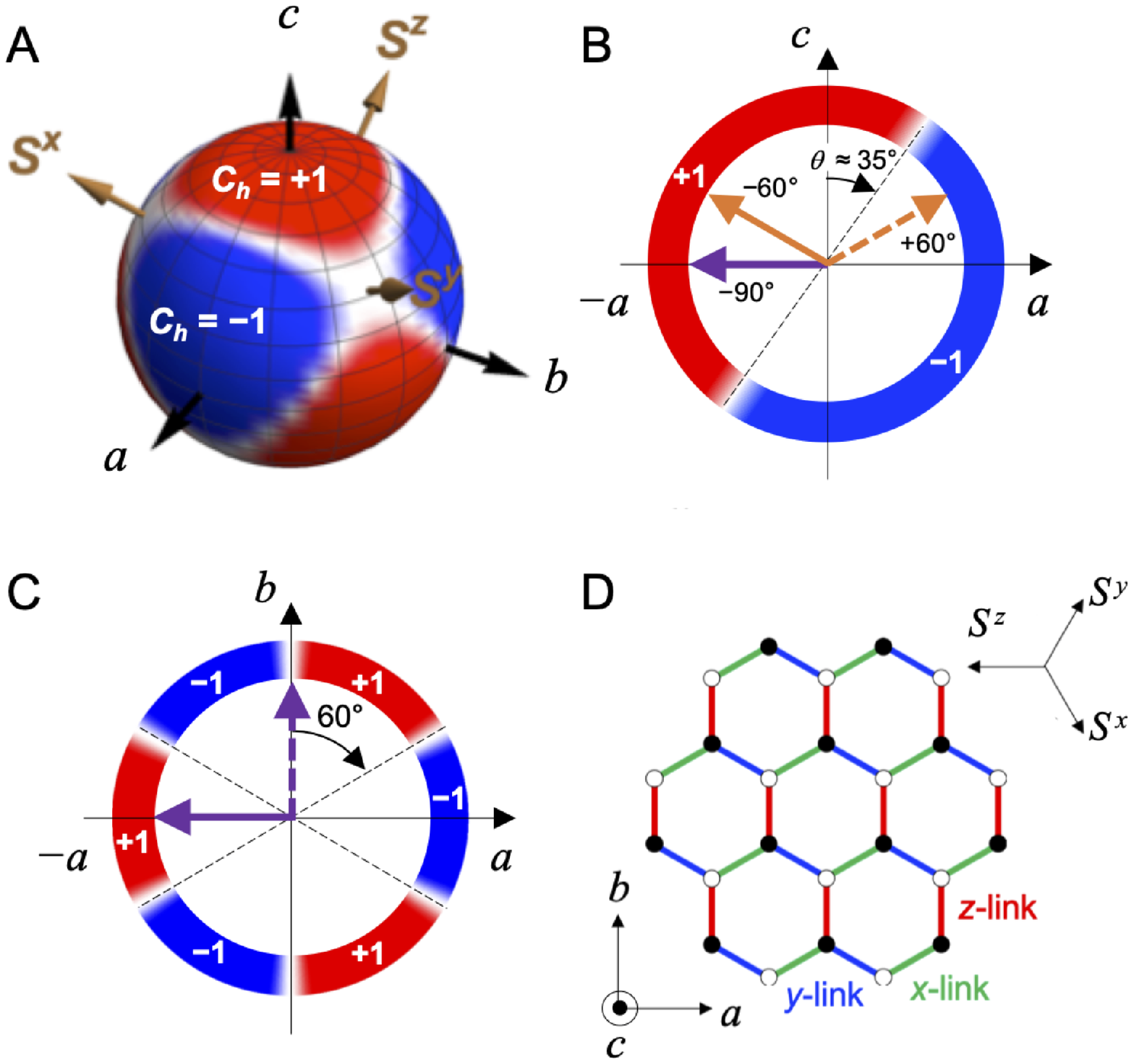}
		\caption{
({\bf A}) Theoretical results for the field-angular variation of Chern number $C_h$, which is $\pm1$,  for the pure Kitaev model with respect to the spin and crystal axes.   ({\bf B}) Variation of $C_h$ within the $ac$ plane.  Orange and orange dashed arrows indicate the field direction at $\theta=-60^{\circ}$ and $+60^{\circ}$ from the $c$ axis, respectively.   Purple arrow indicates the field direction along $-a$ axis ($\theta=-90^{\circ})$.   ({\bf C}) Variation of $C_h$ within the $ab$ plane.  Purple and dashed purple arrows indicate the field direction along $-a$ and $b$ axis, respectively. ({\bf D}) Schematic of honeycomb lattice and its directions.  Two triangular sublattices comprise the honeycomb lattice.    Black and white  circles indicate Ru atoms on  different sublattice. All links are labeled as $x$-, $y$- and $z$-links according to their orientations. Spin axes are perpendicular to these links.
		}
	\end{center}
\end{figure}


\newpage



\bigskip
\bigskip


\begin{thebibliography}{99}
	
	\bibitem{Balents10}
	L. Balents, 
	``Spin liquids in frustrated magnets." 
	Nature {\bf 464}, 199-208  (2010). 
	
	\bibitem{Kitaev06}
	A. Kitaev, 
	``Anyons in a exactly solved model and beyond." 
	Ann. Phys. {\bf 321}, 2-111 (2006). 
	
	\bibitem{Read00}
	N. Read and D. Green, 
	``Paired states of fermions in two dimensions with breaking of parity and time-reversal symmetries and the fractional quantum Hall effect." 
	Phys. Rev. B {\bf 61}, 10267-10297 (2000). 
	
	
	\bibitem{MooreRead91}
	G. Moore and N. Read, 
	``Nonabelions in the fractional quantum Hall effect." 
	Nucl. Phys. B {\bf 360}, 362-396 (1991).
	
	
	\bibitem{Banerjee18}
	M. Banerjee, M. Heiblum, V. Umansky, D.E. Feldman, Y. Oreg, and A. Stern, 
	``Observation of half-integer thermal Hall conductance." 
	Nature {\bf 559}, 205-211 (2018).
	
	\bibitem{Kasahara}
	Y. Kasahara, T. Ohnishi, Y. Mizukami, O. Tanaka, Sixiao Ma, K. Sugii, N. Kurita, H. Tanaka, J. Nasu, Y. Motome, T. Shibauchi, and Y. Matsuda, 
	``Majorana quantization and half-integer thermal quantum Hall effect in a Kitaev spin liquid." 
	Nature {\bf 559}, 227-231 (2018).
	
	\bibitem{Plumb14}
	K. W. Plumb, J. P. Clancy, L. J. Sandilands, V. V. Shankar, Y. F. Hu, K. S. Burch, H.-Y. Kee, and Y.-J. Kim, 
	``$\alpha$-RuCl$_3$: A spin-orbit assisted Mott insulator on a honeycomb lattice." 
	Phys. Rev. B {\bf 90}, 041112(R) (2014).
	
	\bibitem{Jackeli09}
	G. Jackeli and G. Khaliullin, 
	``Mott Insulators in the Strong Spin-Orbit Coupling Limit: From Heisenberg to a Quantum Compass and Kitaev Models." 
	Phys. Rev. Lett. {\bf 102}, 017205 (2009). 
	
	\bibitem{Trebst17}
	S. Trebst, 
	``Kitaev Materials."
	Preprint at http://arxiv.org/abs/1701.07056.
	
	\bibitem{Takagi19}		
	H. Takagi, T. Takayama, G. Jackeli, G. Khaliullin, and  S. E. Nagler, 
	``Concept and realization of Kitaev quantum spin liquids." 
	Nat. Rev. Phys. {\bf 1}, 264-280 (2019).
	
	\bibitem{Kim15}
	H.-S. Kim, V. Vijay Shankar, A. Catuneanu, and H.-Y. Kee, 
	``Kitaev magnetism in honeycomb RuCl$_3$ with intermediate spin-orbit coupling." 
	Phys. Rev. B {\bf 91}, 241110(R) (2015). 
	
	\bibitem{Rau14}
	J. G. Rau, E. K.-H. Lee, and H.-Y. Kee, 
	``Generic Spin Model for the Honeycomb Iridates beyond the Kitaev Limit." 
	Phys. Rev. Lett. {\bf 112}, 077204 (2014).
	
	\bibitem{Winter17}
	S. M. Winter, A. A. Tsirlin, M. Daghofer, J. van den Brink, Y. Singh, P. Gegenwart, and R. Valent\'{i}, 
	``Models and materials for generalized Kitaev magnetism." 
	J. Phys.: Condens. Matter {\bf 29}, 493002 (2017).	
	
	
	\bibitem{Johnson15} 
	R. D. Johnson, S. C. Williams, A. A. Haghighirad, J. Singleton, V. Zapf, P. Manuel, I. I. Mazin, Y. Li, H. O. Jeschke, R. Valent\'{i}, and R. Coldea, 
	``Monoclinic crystal structure of $\alpha$-RuCl$_3$ and the zigzag antiferromagnetic ground state."
	Phys. Rev. B {\bf 92}, 235119 (2015).
	
	\bibitem{Do17}
	S.-H. Do, S.-Y. Park, J. Yoshitake, J. Nasu, Y. Motome, Y. S. Kwon, D. T.  Adroja, D. J. Voneshen, K. Kim,  T.-H. Jang, J.-H. Park, K.-Y. Choi, and S. Ji,
	``Majorana fermions in the Kitaev quantum spin system  $\alpha$-RuCl$_3$."
	Nat. Phys,  {\bf 13}, 1079-1084  (2017).
	
	\bibitem{Widmann19}
	S. Widmann, V. Tsurkan, D. A. Prishchenko, V. G. Mazurenko, A. A. Tsirlin, and A. Loidl, 
	``Thermodynamic evidence of fractionalized excitations in  $\alpha$-RuCl$_3$." 
	Phys. Rev. B {\bf 99}, 094415 (2019).
	
	\bibitem{Sandilands15}
	L. J. Sandilands, Y. Tian, K. W. Plumb, Y.-J. Kim, and K. S. Burch, 
	``Scattering Continuum and Possible Fractionalized Excitations in $\alpha$-RuCl$_3$." 
	Phys. Rev. Lett. {\bf 114}, 147201 (2015).
	
	\bibitem{Nasu16} 
	J. Nasu, J. Knolle, D. L. Kovrizhin, Y. Motome, and R. M\"ossner, 
	``Fermionic response from fractionalization in an insulating two-dimensional magnet." 
	Nat. Phys. {\bf 12}, 912-915 (2016).
	
	\bibitem{Banerjee16}
	A. Banerjee, C. A. Bridges, J.-Q. Yan, A. A. Aczel, L. Li, M. B. Stone, G. E. Granoroth, M. D. Lumsden, Y. Yiu, J. Knolle, S. Bhattacharjee, D. L. Kovrizhin, R. M\"ossner, D. A. Tennant, D. G. Mandrus, and S. E. Nagler, 
	``Proximate Kitaev quantum spin liquid behaviour in a honeycomb magnet." 
	Nat. Mater. {\bf 15}, 733-740 (2016). 
	
	\bibitem{Yoshitake16}
	J. Yoshitake, J. Nasu, and Y. Motome, 
	``Fractional Spin Fluctuations as a Precursor of Quantum Spin Liquids: Majorana Dynamical Mean-Field Study for the Kitaev Model." 
	Phys. Rev. Lett. {\bf 117}, 157203 (2016). 
	
	\bibitem{Banerjee17} 
	A. Banerjee, J. Yan, J. Knolle,  C. A. Bridges, M. B. Stone, M. D. Lumsden, D. G. Mandrus, D. A. Tennant, R. Moessner, and S. E. Nagler, 
	``Neutron scattering in the proximate quantum spin liquid $\alpha$-RuCl$_3$." 
	Science  {\bf 356}, 1055-1059 (2017).
	
	\bibitem{Balz19}
	C. Balz, P. Lampen-Kelly, A. Banerjee, J. Yan, Z. Lu, X. Hu, S. M. Yadav, Y. Takano, Y. Liu, D. A. Tennat, M. D. Lumsden, D. Mandrus, and S. E. Nagler, 
	``Magnons, fractional excitations, and field-induced transitions in $\alpha$-RuCl$_3$."
	Preprint at http://arxiv.org/abs/1903.00056 (2019). 
	
	\bibitem{Nasu17}
	J. Nasu, J. Yoshitake, and Y. Motome, 
	``Thermal Transport in the Kitaev Model."
	Phys. Rev. Lett. {\bf 119}, 127204 (2017).
	
	\bibitem{Rosch18}
	Y. Vinkler-Aviv and A. Rosch, 
	``Approximately Quantized Thermal Hall Effect of Chiral Liquids Coupled to Phonons." 
	Phys. Rev. X {\bf 8}, 031032 (2018).
	
	\bibitem{Balents18}
	M. Ye, G. B. Hal\'asz, L. Savary, and L. Balents, 
	``Quantization of the Thermal Hall Conductivity at Small Hall Angles." 
	Phys. Rev. Lett. {\bf 121}, 147201 (2018).
	
	\bibitem{Kasahara18}
	Y. Kasahara, K. Sugii, T. Ohnishi, M. Shimozawa, M. Yamashita, N. Kurita, H. Tanaka, J. Nasu, Y. Motome, T. Shibauchi, and Y. Matsuda, 
	``Unusual Thermal Hall Effect in a Kitaev Spin Liquid Candidate $\alpha$-RuCl$_3$."
	Phys. Rev. Lett. {\bf 120}, 217205 (2018).
	
	
	\bibitem{Hentrich19} 
	R. Hentrich, M. Roslova, A. Isaeva, T. Doert, W. Brenig, B. B\"uchner, and C. Hess, 
	``Large Thermal Hall Effect in $\alpha$-RuCl$_3$: Evidence for Heat Transport by Kitaev-Heisenberg Paramagnons." 
	Phys. Rev. B {\bf 99}, 085136 (2019).
	
	
	
	
	
	\bibitem{Modic19}
	K. A. Modic, R. D. McDonald, J. P. C. Ruff, M. D. Bachmann, Y. Lai, J. C. Palmstrom, D. Graf, M. Chan, F. F. Balakirev, J. B. Betts, G. S. Boebinger, M. Schmidt, D. A. Sokolov, P. J. W. Moll, B. J. Ramshaw, and A. Shekhter, 
	``Scale-invariance of a Spin Liquid in High Magnetic Fields." 
	Preprint at http://arxiv.org/abs/1901.09245. 
	
	\bibitem{Moon19}
	A. Go, J. Jung, and E.-G. Moon, 
	``Vestiges of Topological Phase Transitions in Kitaev Quantum Spin Liquids.'' 
	Phys. Rev. Lett. {\bf 122}, 147203 (2019).
	
	
	\bibitem{Onose10}
	Y. Onose, T. Ideue, H. Katsura, Y. Shiomi, N. Nagaosa, and Y. Tokura, 
	``Observation of the Magnon Hall Effect." 
	Science {\bf 329}, 297-299 (2010). 
	
	\bibitem{Cookmeyer18}
	J. Cookmeyer and J. E. Moore, 
	``Spin-wave analysis of the low-temperature thermal Hall effect in the candidate Kitaev spin liquid $\alpha$-RuCl$_3$." 
	Phys. Rev. B {\bf 98}, 060412(R) (2018). 
	
	\bibitem{SM}
	Materials, methods, and additional data are available as supplementary materials. 
		
\end{thebibliography}
\end{document}